\documentclass[showpacs,preprintnumbers,amsmath,amssymb,prb,superscriptaddress]{revtex4-1}
\usepackage{graphicx}

\usepackage{color}

\begin{document}

\title{
Resonant inelastic x-ray scattering study of entangled spin-orbital excitations in superconducting PrFeAsO$_{0.7}$}

\author{T.~Nomura}
\email{nomurat@spring8.or.jp}
\affiliation{National Institutes for Quantum and Radiological Science and Technology (QST), SPring-8, 1-1-1 Kouto, Sayo, Hyogo 679-5148, Japan}

\author{Y.~Harada}
\affiliation{Institute for Solid State Physics (ISSP), University of Tokyo, Kashiwanoha, Kashiwa, Chiba 277-8581, Japan}

\author{H.~Niwa}
\affiliation{Division of Physics, Faculty of Pure and Applied Sciences, University of Tsukuba, Tsukuba, Ibaraki 305-8571, Japan}

\author{K.~Ishii}
\affiliation{National Institutes for Quantum and Radiological Science and Technology (QST), SPring-8, 1-1-1 Kouto, Sayo, Hyogo 679-5148, Japan}

\author{M.~Ishikado}
\affiliation{Research Center for Neutron Science and Technology, Comprehensive Research Organization for Science and Society (CROSS), Tokai, Ibaraki
319-1106, Japan}

\author{S.~Shamoto}
\affiliation{Advanced Science Research Center, Japan Atomic Energy Agency (JAEA), Tokai, Ibaraki 319-1195, Japan}

\author{I.~Jarrige}
\affiliation{National Synchrotron Light Source II, Brookhaven National Laboratory, Upton, New York, 11973, USA}

\date{\today}

\begin{abstract}
Low-energy electron excitation spectra were measured on a single crystal of a typical iron-based superconductor PrFeAsO$_{0.7}$ using resonant inelastic X-ray scattering (RIXS) at the Fe-$L_3$ edge. Characteristic RIXS features are clearly observed around 0.5, 1-1.5 and 2-3~eV energy losses. These excitations are analyzed microscopically with theoretical calculations using a 22-orbital model derived from first-principles electronic structure calculation. Based on the agreement with the experiment, the RIXS features are assigned to Fe-$d$ orbital excitations which, at low energies, are accompanied by spin flipping and dominated by Fe $d_{yz}$ and $d_{xz}$ orbital characters. 
Furthermore, our calculations suggest dispersive momentum dependence of the RIXS excitations below 0.5 eV, 
and predict remarkable splitting and merging of the lower-energy excitations in momentum space. 
Those excitations, which were not observed in the present experiment, highlight the potential of RIXS with an improved energy resolution to unravel new details 
of the electronic structure of the iron-based superconductors.
\end{abstract}


\maketitle


\section{Introduction}

Since the discovery of high-$T_{\rm c}$ superconductivity in iron pnictides~\cite{Kamihara2008}, 
extensive experimental and theoretical efforts have been devoted to elucidate the underlying mechanism of this intriguing physical phenomenon~\cite{Hosono2015}. 
The key ingredient for the pairing mechanism is the attractive interaction between electrons forming Cooper pairs~\cite{Scalapino2012}. 
Various microscopic origins have been proposed for the pairing attraction in iron pnictides so far. 
In promising pairing scenarios, electronic elementary excitations such as 
antiferromagnetic fluctuations~\cite{Singh2008,Kuroki2008} and orbital fluctuations~\cite{Kontani2010,Yanagi2010} are proposed to serve the role of mediator of the pairing. 
The fact that superconductivity in most of the iron-based superconductors emerges in the proximity of an antiferromagnetic transition or a structural transition gives further 
credence to these scenarios. It also shows that a detailed knowledge of the elementary excitations can be an essential clue to discuss pairing mechanisms. 

Recently, resonant inelastic x-ray scattering (RIXS) has emerged as a powerful technique 
to observe various elementary excitations in solids~\cite{Ament2011,Ishii2013}. 
Particularly, RIXS at the transition-metal absorption edges is useful 
to study the dynamics of strongly correlated $d$ electrons in transition-metal compounds. 
The type of elementary excitations that can be observed depends on the utilized absorption edge. 
While $K$-edge RIXS is appropriate for studying charge dynamics, 
$L$-edge RIXS enables studies of not only charge-orbital dynamics 
but also spin dynamics. This contrast can be attributed to the difference in the RIXS intermediate state. 
In $K$-edge RIXS, when the electric-dipole transition dominates the resonant transition at the main edge, 
an inner-shell $1s$ electron is promoted to an empty $4p$ state.
Transition-metal $d$ electrons near the Fermi energy ($E_F$) 
are excited to screen the $1s$ hole with a spin-independent isotropic potential. 
On the other hand, in $L$-edge RIXS, the $2p$ electrons are promoted to transition-metal $d$ bands. 
Since the $2p$ states split into $j=1/2$ doublets and $j=3/2$ quartets 
due to the strong spin-orbit coupling, the promoted electrons are polarized in spin in general. 
Therefore magnetic excitations can be induced within the transition-metal $d$ bands~\cite{Ament2009,Haverkort2010,Igarashi2012}. 
When both the incoming and outgoing x-rays are linearly polarized as in usual RIXS experiments, 
the orbital angular momentum of the $d$-electron system can also change in the final state, 
since the total angular momentum of spin and orbital is conserved.
Thus we should note that $L$-edge RIXS can detect electronic excitations which 
$K$-edge RIXS is insensitive to, e.g., single-spin flip excitations and off-diagonal orbital excitations 
(Throughout the present article, we refer to excitations where the orbital states of the excited electron 
and hole are the same [different from each other] in the final state, as {\it diagonal} [{\it off-diagonal}] orbital excitations). 
$L$-edge RIXS has indeed been widely applied to a number of copper oxides, 
resulting in successful observations of magnetic and orbital excitations
~\cite{Braicovich2010,Guarise2010,LeTacon2011,Sala2011,Schlappa2012,Dean2012,Ishii2014,Lee2014}. 
On the other hand, only fewer $L$-edge RIXS studies have been performed for iron pnictides 
or chalcogenides so far, to our knowledge~\cite{Yang2009,Hancock2010,Zhou2012,Monney2013}. 

In the iron-based superconductors, first-principles electronic structure calculations strongly indicate 
that each of the Fe-$d$ orbitals occupies a significant part of density of states (DOS) near $E_F$~\cite{Singh2008,Ishibashi2008}. 
It is therefore expected that the nature of the low-energy electronic excitations is in principle 
quite different from that of the high-$T_c$ cuprate superconductors, in which only the Cu-$d_{x^2-y^2}$ 
orbital is dominant near $E_F$. In particular, it was suggested that orbital fluctuations involving the Fe-$d_{yz}$ and $d_{xz}$ orbitals
could play a role in inducing the pairing in the iron pnictides~\cite{Kontani2010}. 
This calls for a study of the low-energy orbital excitations in these materials using $L$-edge RIXS. 

Previously, Jarrige, Gretarsson and collaborators performed $K$-edge RIXS of a typical superconducting iron-pnictide PrFeAsO$_{0.7}$~\cite{Jarrige2012} 
and insulating iron chalcogenide K$_{0.83}$Fe$_{1.53}$Se$_2$~\cite{Gretarsson2015}. In both cases, the momentum dependence of the RIXS spectra could be explained
by assuming that the Coulomb interaction between Fe-$d$ electrons 
should be moderately strong, $U \approx 2.4$ - 3 eV. 
Consistency between the experiment and {\it ab initio} calculations suggested 
that the excitation spectra are dominated by orbital excitations with the Fe-$d_{yz}$ and $d_{xz}$ 
character without any spin flip. 

In the present paper, we report $L$-edge RIXS study for a typical iron-based superconductor, PrFeAsO$_{0.7}$. 
We found clearly characteristic RIXS features around excitation energies of 0.5, 1-1.5 and 2-3 eV. 
To interpret these features, we carried out a theoretical study based 
on an electronic band structure calculation. The experimental features are well captured 
by assuming $U \approx 3$ eV as in the previous $K$-edge RIXS studies ~\cite{Jarrige2012,Gretarsson2015}. 
Based on the agreement with theory, we are able to assign those features to orbital excitations 
among Fe-$d$ orbitals at a microscopic level. 
Furthermore, our calculation suggests that single-magnon excitations 
and spin-flipped orbital excitations should appear at excitation energies below 0.5 eV, 
which are dispersive with respect to x-ray momentum transfer. 
These excitations were not observed in the present experiment, likely due to the limited energy resolution and excitation damping.
Remarkable splitting and merging of the lower-energy RIXS peaks in momentum space 
are predicted, which have not been experimentally observed so far. 

\section{Experiment}
The RIXS spectra were collected at the beamline BL07SU and the x-ray emission spectrometer HORNET \cite{Harada2012} at SPring-8, Japan. The total energy resolution for the RIXS measurements was 
$\sim$230 meV at the Fe-$L_3$ edge, which corresponds to a resolving power 
$E/\Delta E \approx 3000$. 
The x-ray absorption spectroscopy (XAS) spectra were measured in the total-fluorescence-yield (TFY) mode. Single crystals of PrFeAsO$_{0.7}$ were grown by a high-pressure synthesis method using a belt-type anvil apparatus described in Ref. \onlinecite{Ishikado2010}. The sample belongs to the so-called 1111 family, which crystallizes in the ZrCuSiAs-type structure, in the tetragonal space group $P4/nmm$. 
In this sample, the electron doping of 0.6 is optimal and yields a $T_c$ of 42~K. 
The scattering geometry was chosen to minimize the intensity of the elastic peak. 
The spectrometer arm was placed at 90 degrees from the incident beam in the horizontal scattering plane, 
and incoming x-rays were always horizontally polarized, i.e., $\pi$-polarized. 
All data were taken at room temperature. 

\begin{figure}[htb]
\includegraphics[width=0.8\columnwidth]{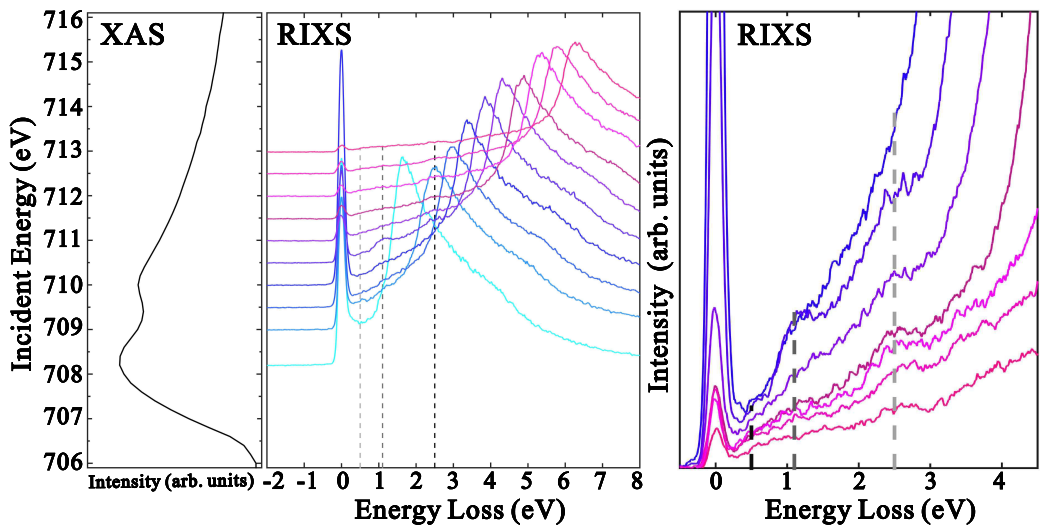}
\caption{(Color online)
Fe-$L_{3}$ XAS spectrum measured in the total fluorescence yield mode (left panel) and RIXS spectra measured for a few incident energies across the edge (center panel) on PrFeAsO$_{0.7}$. 
The vertical offset of the RIXS spectra matches their respective incident energies along the XAS spectrum energy axis. The vertical dashed lines indicate the energy loss of the Raman-like features. 
X-ray momentum transfer was ${\bf Q} \approx (0,0,1.4\pi)$. 
Close-up of the low-energy portion of the RIXS spectra of PrFeAsO$_{0.7}$, with no vertical offset (right panel).}
\label{fig1}
\end{figure}

RIXS spectra measured on PrFeAsO$_{0.7}$ at a few incident x-ray energies $E_{inc}$ across the $L_{3}$ edge are shown in Figure~\ref{fig1}. The vertical offset of the RIXS spectra is scaled to the energy axis of the XAS spectrum. The RIXS spectra display a salient feature around 1.5~eV energy loss at $E_{inc}$=708~eV which tracks the incident energy up to $\sim$ 6.5~eV at $E_{inc}$=713~eV. 
This behavior is typical of fluorescence, and is observed not only in iron pnictides but also in Fe metal and $\alpha$-Fe$_2$O$_3$~\cite{Yang2009}. 
We assign this peak to the $L\alpha_{1}$ emission line, which corresponds to the $3d_{5/2} \rightarrow 2p_{3/2}$ fluorescent decay.
As the fluorescence disperses to higher energy losses, weak features appear in its low-energy loss tail, and remain at fixed energy loss upon increase in the incident energy, around 0.5, 1 and 2.5~eV. These RIXS features are related to charge excitations, as discussed in the next section. 
We note that while the RIXS features are very weak, such a clear observation of Fe $L$-edge RIXS excitations in a 1111 Fe-based superconductor had previously not been reported 
in the literature to the knowledge of these authors. A close-up of the low-energy portion of the RIXS spectra, with no vertical offset, is shown in the right panel of Figure~\ref{fig1}. 
The presence of well-defined Raman-like features in the PrFeAsO$_{0.7}$ data can be confirmed.

We note that the lineshape and energy loss of the $L$-edge RIXS excitations are sharply different from the $K$-edge data \cite{Jarrige2012,Gretarsson2015}. As discussed in the next section, this is not unexpected since the spectral weight in $K$-edge RIXS is dominated by orbital-diagonal transitions, 
whereas the $L$-edge spectral weight mostly arises from orbital off-diagonal transitions.

\section{Theoretical Calculation}

\subsection{Theoretical framework}

\begin{figure}[htb]
\includegraphics[width=0.6\columnwidth]{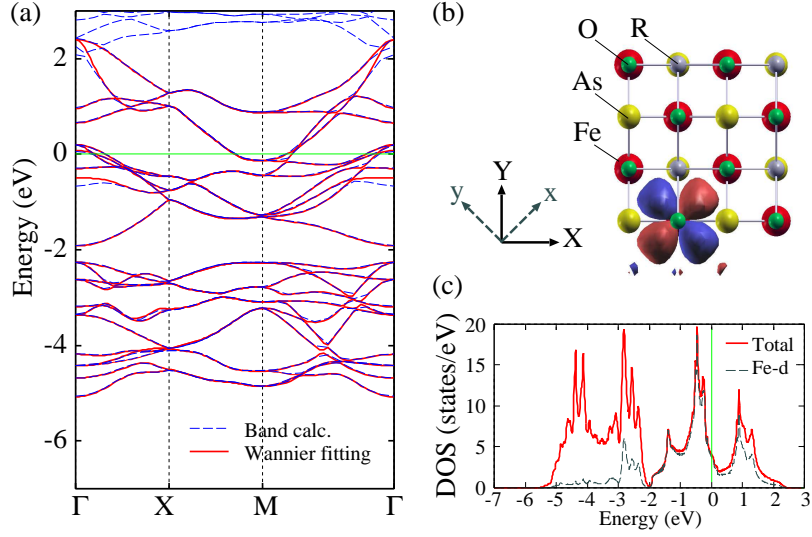}
\caption{(Color online)
(a) Result of band structure calculation and fitting by the 22 MLWS's near $E_F$, 
(b) Crystal structure viewed from the $Z$ direction and the Fe-$d_{x^2-y^2}$ MLWS. 
R represents a rare-earth atom. 
(c) Density of states (DOS) based on MLWS's. In (a) and (c), $E_F$ is set to zero.}
\label{fig2}
\end{figure}

To analyze the observed RIXS spectra, we start with a first-principles electronic 
structure of LaFeAsO, using the {\sc WIEN}2k code~\cite{Blaha2015} (See Fig.~\ref{fig2}(a)). 
We may assume that the Pr system possesses a similar electronic structure near $E_F$, 
since Pr-$f$ electrons will almost completely localize. 
We may also assume that oxygen vacancies in actual Pr samples 
only slightly change the electronic band structure and $E_F$, 
judging from the results of x-ray absorption and emission spectroscopy (XAS and XES)~\cite{Freelon2010}. 
From the calculated electron bands, we construct an effective 22-band model 
near $E_F$, using the {\sc wannier90} code~\cite{Mostofi2008}, 
where ten Fe-$d$, six As-$p$ and six O-$p$ maximally localized Wannier states (MLWS) are included 
(Note there are two Fe, two As and two O atoms in the unit cell. See Fig.~\ref{fig2}(b)). 
Throughout the present study, we express MLWS's by using the local coordinates $x$, $y$ and $z$ 
as in Fig.~\ref{fig2}(b), while x-ray momenta and scattering geometry below shall be specified 
by the global coordinates $X$, $Y$ and $Z$. The crystalline [100], [010] and [001] axes correspond 
to the $X$, $Y$ and $Z (\parallel z)$ axes, respectively. 
As shown in Fig.~\ref{fig2}(c), DOS near $E_F$ is dominated by the Fe-$d$ MLWS's, consistent 
with previous first-principles band calculations~\cite{Singh2008,Ishibashi2008}.
Taking the on-site Coulomb interaction at Fe sites into account, 
we determine the antiferromagnetic (AF) ground state within the Hartree-Fock approximation (HFA), 
where we assume the AF ordering wave vector ${\bf Q}_{AF}=(\pi,\pi,\pi)$ and magnetic moments pointing 
along the [110] direction, as observed in experiments~\cite{DeLaCruz2008}.
We take $U=3$ eV, $U'=0.6U$, $J=J'=0.2U$, as in the previous $K$-edge RIXS study~\cite{Jarrige2012}. 
These Coulomb integrals agree well with the evaluation from ab-initio calculations~\cite{Miyake2010}. 
Calculated spin moments are $m_{xy} = 0.68$,  $m_{yz} = m_{xz} = 0.51$, $m_{x^2-y^2} = 0.28$, 
$m_{3z^2-r^2} = 0.59$, in units of $\mu_B$. 

Recently, one of the authors has developed a theoretical framework of $L$-edge RIXS 
based on a perturbation theory~\cite{Nomura2015} with respect to electron-electron interaction, 
where the AF ground state is determined within HFA 
and electron correlations are dealt with by the random-phase approximation (RPA). 
Here we present the formula for RIXS intensity without derivation 
(see Ref.~\onlinecite{Nomura2015} for details of its derivation): 
\begin{eqnarray}
W(q, {\bf e}; q', {\bf e}') &=& 2 \pi \sum_{{\bf k}_1} \sum_{a_1 a_2} 
n_{a_1}({\bf k}_1) [ 1-n_{a_2}({\bf k}_1 + {\bf Q}) ]
\delta (\Omega + E_{a_1}({\bf k}_1)- E_{a_2}({\bf k}_1 + {\bf Q})) \nonumber \\
&& \times \biggl| \sum_i^{\rm t.m.u.} \sum_{\ell\ell'}^{@{\bf r}_i} \sum_{\sigma\sigma'} 
\sum_{jj'} \sum_{m=-j}^{j} \sum_{m'=-j'}^{j'} w_{\ell\sigma, jm}({\bf r}_i; {\bf q}, {\bf e}) 
w_{\ell'\sigma', j'm'}^*({\bf r}_i; {\bf q}', {\bf e}')  \nonumber \\
&& \times F_{\ell\sigma jm, \ell'\sigma' j'm'; a_1, a_2}({\bf r}_i; {\bf k}_1; q, q') \biggr|^2. 
\label{Eq:W}
\end{eqnarray}
where $q=(\omega, {\bf q})$ and $q=(\omega', {\bf q}')$ are the four-momenta 
of incoming and outgoing x-rays, respectively. 
The energy loss and momentum change of the x-ray are $Q= (\Omega, {\bf Q}) 
= (\omega-\omega', {\bf q}-{\bf q}')$. 
${\bf e}$ and ${\bf e}$' are the polarization vectors of incoming and outgoing x-rays, respectively.
$E_a({\bf k})$ and $n_a({\bf k})$ are the band energy and electron occupation number 
at momentum {\bf k} on band $a$ within HFA, respectively. 
In the present calculation, we have 88 bands ($1 \leq a \leq 88$) as a consequence from band-folding and spin degeneracy.
In derivation of Eq.~($\ref{Eq:W}$), it is assumed that only a single electron-hole pair is left in the final state. 
In numerical calculation of Eq.~(\ref{Eq:W}), we use the Lorentzian form for the $\delta$-function:
\begin{equation}
\delta (z) \approx \frac{\gamma}{\pi(z^2 + \gamma^2)}, 
\end{equation}
where $\gamma$ is a broadening factor of calculated spectra 
(Hereafter, unless we notify, we take $\gamma = 0.02$ eV). 
The Fe-$2p$ states are specified by the total angular momentum quantum numbers $j$ and $m$, 
where $j=1/2, 3/2$ and $m = -j, ... ,+j$. 
$w_{\ell\sigma, jm}({\bf r}_i; {\bf q}, {\bf e})$ is the matrix elements of electric-dipole transition 
from Fe-$2p_{jm}$ to Fe-$d_{\ell\sigma}$ state at iron site $i$, 
where $\ell = xy$, $yz$, $xz$, $x^2-y^2$, $3z^2-r^2$ and $\sigma$ is spin index. 
t.m.u. in summation in $i$ means summing in $i$ only over the transition-metal sites in the unit cell.  
$@{\bf r}_i$ in summation in $\ell$ and $\ell'$ means that $d$ orbitals $\ell$ and $\ell'$ 
should reside on transition-metal site $i$. 
$F_{\ell \sigma jm, \zeta' \ell'\sigma'; a_1, a_2}({\bf r}_i; {\bf k}_1; q, q')$ 
is a scattering vertex function, which is the sum of three parts: 
\begin{eqnarray}
F_{\ell\sigma jm, \ell'\sigma' j'm'; a_1, a_2}({\bf r}_i; {\bf k}_1; q, q') 
&=& F^{(0)}_{\ell\sigma jm, \ell'\sigma' j'm'; a_1, a_2}({\bf r}_i; {\bf k}_1; q, q') \nonumber \\
&& - \sum_{\ell_1\ell_2} \sum_{\sigma_1\sigma_2} 
u_{\ell_2\sigma_2, a_2}^*({\bf k}_1 + {\bf Q}) u_{\ell_1\sigma_1, a_1}({\bf k}_1) 
[ F^{(p)}_{\ell\sigma jm, \ell'\sigma' j'm'; \ell_1\sigma_1, \ell_2\sigma_2}({\bf r}_i; q, q') \nonumber \\
&& + F^{(d)}_{\ell\sigma jm, \ell'\sigma' j'm'; \ell_1\sigma_1, \ell_2\sigma_2}({\bf r}_i; q, q') ]. 
\label{Eq:F}
\end{eqnarray}

The first part is given by 
\begin{equation}
F^{(0)}_{ \ell\sigma jm, \ell'\sigma' j'm'; a_1, a_2}({\bf r}_i; {\bf k}_1; q, q') = \delta_{jj'}\delta_{mm'} 
\frac{u_{\ell \sigma,a_2}^*({\bf k}_1+{\bf Q}) u_{\ell'\sigma', a_1}({\bf k}_1)}
{\omega + \tilde{\varepsilon}_{2p_j}({\bf r}_i) - E_{a_2}({\bf k}_1+{\bf Q})}, 
\label{eq:F0}
\end{equation}
where $\ell\sigma$ means $d_{\ell}$ state with spin $\sigma$ at iron site $i$, 
and $u_{\ell\sigma, a}({\bf k})$ is the diagonalization matrix of the Hamiltonian in HFA. 
$\tilde{\varepsilon}_{2p_j}({\bf r}_i) \equiv \varepsilon_{2p_j}({\bf r}_i) + i \Gamma_{2p}$ 
is the energy of $2p$ states with a damping rate $\Gamma_{2p}$. 
For the present work, we take $\varepsilon_{2p_{1/2}}({\bf r}_i) = - 722.2$ eV 
and $\varepsilon_{2p_{3/2}}({\bf r}_i) = - 709.15$ eV with respect to $E_F$ and $\Gamma_{2p} = 0.3$ eV. 
This part describes the most simple lowest-order RIXS process: 
a $2p$ electron is promoted to an empty Fe-$d$ state above $E_F$ by absorbing incident x-ray 
($\omega, {\bf q}, {\bf e}$), and then an Fe-$d$ electron below $E_F$ decays into the empty $2p$ state, 
emitting x-ray ($\omega', {\bf q}', {\bf e}'$). 
This process is a simple inter-band transition of the 0th order with respect 
to the electron-electron Coulomb interaction. 

The second part is the indirect process where Fe-$d$ electrons near $E_F$ are excited 
to screen the created inner-shell $2p$ hole. 
\begin{eqnarray}
F^{(p)}_{ \ell\sigma jm, \ell'\sigma' j'm'; \ell_1\sigma_1, \ell_2\sigma_2}({\bf r}_i; q, q') &=& 
\sum_{\ell_3\ell_4}^{@{\bf r}_i} \sum_{\sigma_3\sigma_4}
V_{2p-d}({\bf r}_i; jm, \ell_3\sigma_3 ; \ell_4\sigma_4, j'm') 
\Lambda_{\ell_2\sigma_2, \ell_4\sigma_4 ; \ell_3\sigma_3, \ell_1\sigma_1}(Q) \nonumber \\
&& \times \sum_a \sum_{{\bf k}} [1-n_a({\bf k})] \nonumber\\
&& \times \frac{u_{\ell\sigma, a}^*({\bf k}) u_{\ell'\sigma', a}({\bf k})}
{[\omega + \tilde{\varepsilon}_{2p_j}({\bf r}_i) - E_a({\bf k})]
[\omega' + \tilde{\varepsilon}_{2p_{j'}}({\bf r}_i) - E_a({\bf k})]}, 
\label{eq:Fp}
\end{eqnarray}
where $V_{2p-d}({\bf r}_i; jm, \ell_3\sigma_3 ; \ell_4\sigma_4, j'm')$ is the inter-orbital Coulomb 
interaction between Fe-$2p$ and Fe-$d$ electrons at iron site $i$, 
and is treated within the Born approximation. 
In the present work, we take the Slater-Condon parameters as $F^0_{pd} = F^2_{pd} = 2$ eV 
to determine $V_{2p-d}({\bf r}_i; jm, \ell_3\sigma_3 ; \ell_4\sigma_4, j'm')$. 
$\Lambda_{\ell_2\sigma_2, \ell_4\sigma_4 ; \ell_3\sigma_3, \ell_1\sigma_1}(Q)$ 
is a vertex function, which describes multiple scattering between Fe-$d$ electrons. 
We calculate $\Lambda_{\ell_2\sigma_2, \ell_4\sigma_4 ; \ell_3\sigma_3, \ell_1\sigma_1}(Q)$ 
within RPA with respect to the Fe-$d$ Coulomb interaction. 

The third part is given by: 
\begin{eqnarray}
F^{(d)}_{ \ell\sigma jm, \ell'\sigma' j'm'; \ell_1\sigma_1, \ell_2\sigma_2}({\bf r}_i; q, q') = 
\delta_{jj'}\delta_{mm'} \sum_{\ell_3\ell_4} \sum_{\sigma_3\sigma_4} 
\Gamma_{\ell_2\sigma_2, \ell_4\sigma_4; \ell_3\sigma_3, \ell_1\sigma_1}(Q)
\sum_{a_3 a_4} \sum_{{\bf k}} [1-n_{a_3}({\bf k}+{\bf Q})] \nonumber \\
\times \frac{u_{\ell\sigma, a_3}^*({\bf k}+{\bf Q}) u_{\ell_3\sigma_3, a_3}({\bf k}+{\bf Q}) 
u_{\ell_4\sigma_4, a_4}^*({\bf k}) u_{\ell'\sigma', a_4}({\bf k})} 
{\omega + \tilde{\varepsilon}_{2p_j}({\bf r}_i) - E_{a_3}({\bf k}+{\bf Q})} \nonumber\\
\times \biggl( \frac{1-n_{a_4}({\bf k})}{\omega' + \tilde{\varepsilon}_{2p_{j'}}({\bf r}_i) - E_{a_4}({\bf k})} 
- \frac{n_{a_4}({\bf k})}{\Omega+E_{a_4}({\bf k}) - E_{a_3}({\bf k}+{\bf Q}) + i\gamma} \biggr), 
\nonumber \\
\label{eq:Fd}
\end{eqnarray}
where $\Gamma_{\ell_2\sigma_2,\ell_4\sigma_4; \ell_3\sigma_3,\ell_1\sigma_1}(Q)$ 
is another vertex function, which we calculate within RPA, 
as for $\Lambda_{\ell_2\sigma_2, \ell_4\sigma_4 ; \ell_3\sigma_3, \ell_1\sigma_1}(Q)$. 
This part contains higher-order processes with respect to the Fe-$d$ Coulomb interaction, 
such as multiple scatterings between excited Fe-$d$ electron and hole in the intermediate state. 

In our theoretical framework, local $dd$ excitations are included in terms of the scattering functions $F^{(p)}$ and $F^{(d)}$, 
and spin-flip (single-magnon) excitations are included mainly in terms of the scattering function $F^{(d)}$. 
$F^{(0)}$ describes simple inter-band transitions. 
For $F^{(0)}$ in Eq.~(\ref{eq:F0}), $\ell\sigma$ and $\ell'\sigma'$ represent the orbital-spin states 
of the electron left above $E_F$ and hole below $E_F$ in the final state, respectively. 
For $F^{(p,d)}$ in Eqs.~(\ref{eq:Fp}) and (\ref{eq:Fd}), $\ell_2\sigma_2$ and $\ell_1\sigma_1$ represent them.
For example, the components of the scattering function $F^{(p,d)}_{ \ell\sigma jm, \ell'\sigma' j'm'; \ell_1\sigma_1, \ell_2\sigma_2}$ 
with $\ell_2 = \ell_1$ [with $\ell_2 \neq \ell_1$] describe diagonal [off-diagonal] orbital excitations. 
In addition, if $\sigma_2 = \sigma_1$ [$\sigma_2 \neq \sigma_1$], we refer to them as spin-conserved [spin-flipped] orbital excitations.

\subsection{Interpretation of experimental data}

In order to compare the experimental and calculated spectra, the scattering geometry of the calculations is matched to the experiment. 
Namely, the scattering plane is parallel to the $XZ$ plane, and $\theta$ and $2\theta$ are respectively set to 45$^{\circ}$ and 90$^{\circ}$. 
We specify the polarization of the incoming [outgoing] x-ray, using a polarization angle 
$\psi$ [$\psi'$], which is the angle between the polarization vector {\bf e} [{\bf e}'] 
and the scattering plane. 
In our present theoretical study, we retain $\psi=0$ as in the experiment, 
which means that the incoming x-rays are always $\pi$-polarized. 
We did not resolve the RIXS spectra with respect to the polarization of outgoing x-rays in the experiment. 
On the other hand, we shall always monitor the dependence on the polarization direction of outgoing x-rays below in the theoretical analysis. 

\begin{figure}[htb]
\includegraphics[width=0.7\columnwidth]{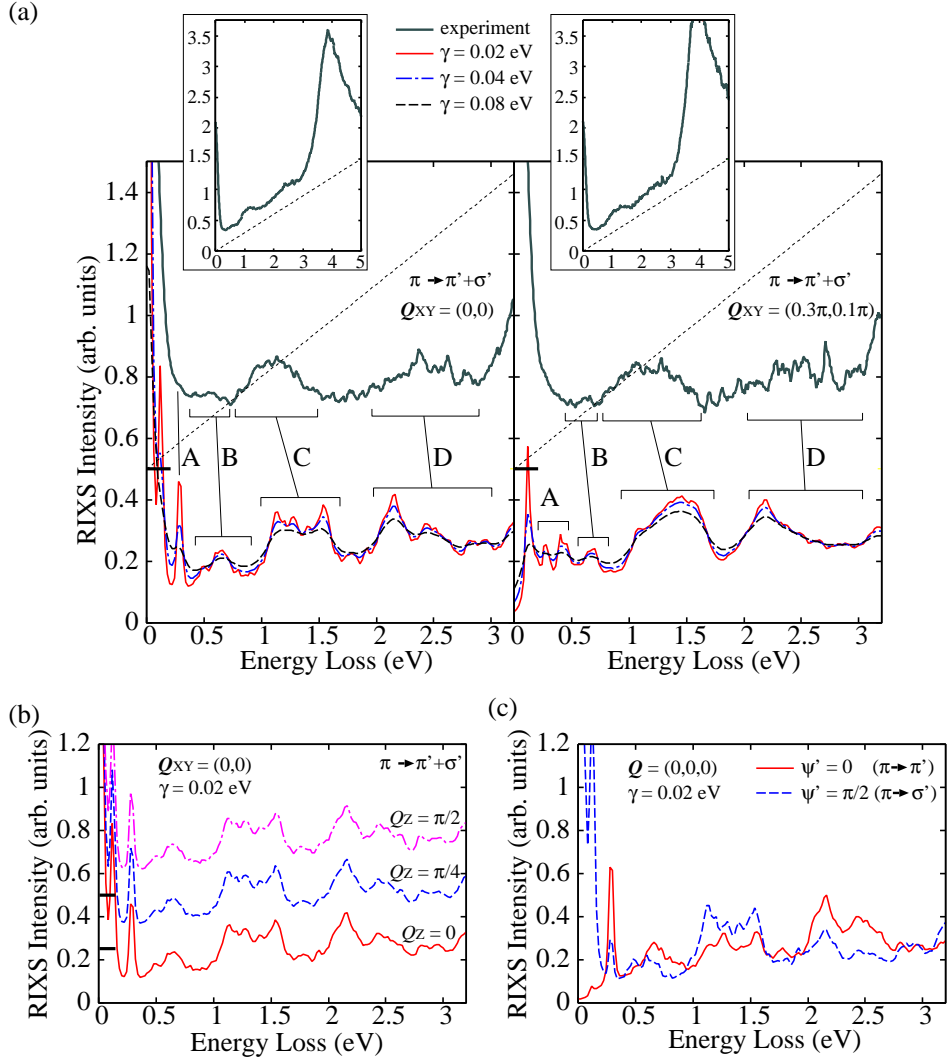}
\caption{(Color online) 
(a) Assigning the observed RIXS features to calculated ones: left and right panels show the results 
at in-plane momentum transfers ${\bf Q}_{XY}=(0,0)$ and $(0.3\pi,0.1\pi)$, respectively. 
The thick horizontal bar and the dotted slope represent the baseline and the subtracted linear background 
of the experimental data, respectively. 
Insets show the subtracted linear background and the experimental curve before subtraction. 
(b) Calculated dependence on out-of-plane momentum transfer $Q_Z$. 
Baselines for $Q_Z = \pi/4$ and $\pi/2$ are vertically shifted for clarity. 
(c) Calculated dependence on the outgoing x-ray polarization. 
In every panel, the incident x-ray is $\pi$-polarized ($\psi=0$), and the incident x-ray energy is set to $\omega = 710.5$ eV. 
In (a) and (b), curves of the spectra averaged for $\psi'= 0$ and $\pi/2$ ($\pi'$- and $\sigma'$-polarizations) are drawn.} 
\label{fig3}
\end{figure}

Typical experimental and calculated RIXS spectra at $\omega=710.5$ eV 
are compared at in-plane momentum transfers ${\bf Q}_{XY}=(0, 0)$ and ${\bf Q}_{XY}=(0.3\pi, 0.1\pi)$ 
in Fig.~\ref{fig3}(a), where, to ease the comparison, 
calculated data are averaged in the polarization of the outgoing x-ray, and a linear background denoted by a dotted slope has been subtracted from the experimental data. 
This background arises from the $L\alpha$ fluorescence signal, which is not taken into account in the calculations. 
The three observed RIXS features B, C and D, which do not drastically depend on in-plane momentum transfer, 
are qualitatively reproduced by the calculation. 
As shown in Fig.~\ref{fig3}(b), the calculation also suggests that the spectra do not depend on out-of-plane momentum transfer $Q_Z$, 
reflecting the two-dimensionality of this compound.

According to our calculation, a sharp low-energy feature A is present around 0.25 eV, 
whose dependence on x-ray momentum transfer we shall discuss closely in the next subsection. 
We consider that this feature A is likely hidden by the tail of the elastic peak in the experimental spectra. 
A possible reason for the low-energy feature A being not visible in the experimental spectra could be related to the broadening of the spectra. 
Experimental broadening can arise both from the resolution limit of the experimental instrument and from the damping of the excitations. 
As seen in Fig.~\ref{fig3}(a), the peak intensity of the low-energy feature A strongly depends on the broadening factor $\gamma$, 
and a broadening of $\gamma=0.08$ eV could explain why the feature was not experimentally observed.

We consider that the feature A is different in nature from the higher-energy features B, C and D.
At the microscopic level, B, C, and D are mainly derived from the zeroth-order processes 
described by the scattering function $F^{(0)}$, where the $2p$ electron is promoted to empty Fe-$d$ levels above $E_F$, 
followed by the decay of an Fe-$d$ electron below $E_F$ into the empty $2p$ state. 
These features correspond to Fe-$d$ interband transitions. 
On the other hand, A is a spin-flipped $dd$ excitation arising mainly from the processes described by the scattering function $F^{(d)}$, 
and does not correspond to the Fe-$d$ interband transitions. 
Since the peak intensity of A depends significantly on the polarization of the outgoing x-ray, as seen Fig.~\ref{fig3}(c), 
the feature A could be distinguished from the background by discriminating the outgoing x-rays in polarization. 

\begin{figure}[htb]
\includegraphics[width=0.5\columnwidth]{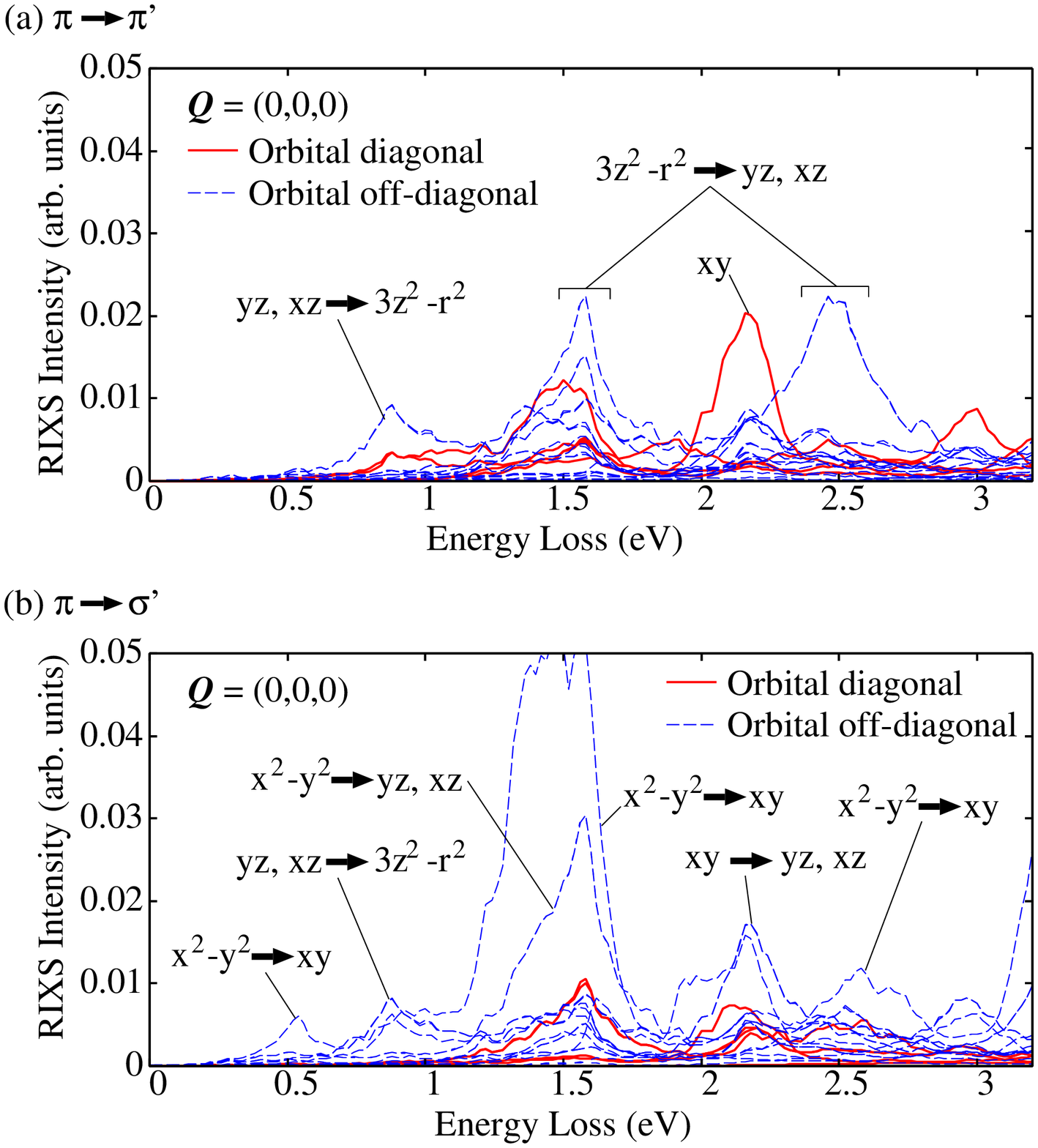}
\caption{(Color online)
Orbital-resolved intensities for charge-orbital channels without any spin flipping. 
X-ray polarizations are (a) $\pi \rightarrow \pi'$ and (b) $\pi \rightarrow \sigma'$. 
The entire 25 processes including both diagonal and off-diagonal orbital excitations 
are plotted in each panel. The thick solid and broken thin curves represent 
diagonal and off-diagonal orbital excitations, respectively.}
\label{fig4}
\end{figure}

To resolve the high-energy features B, C and D into spin-orbital components, 
we project the spectrum onto each spin-orbital excitation process, which can be characterized 
by the spin and orbital characters of the electron-hole pair left in the final state. 
Specifically, we are able to extract a certain process related to a desired final spin-orbital character, 
by constraining the summation over spin and orbital indices of final states 
in the calculation of the intensity $W(q,{\bf e}; q',{\bf e}')$ and the scattering functions~\cite{Nomura2015}. 
To get insights into the orbital nature of B, C and D, we set $F=F^{(0)}$ (simple inter-band transitions) 
and suppress the summation in $\ell$ and $\ell'$ in Eq.~(\ref{Eq:W}). 
We show the orbital-resolved spectra in charge-orbital (i.e., not spin-flipping) channels in Fig.~\ref{fig4}. 
The low-energy sharp structures below 0.5 eV seen in Fig.~\ref{fig3} do not appear, 
since they originate not from the simple inter-band transitions $F^{(0)}$ 
but from many-body correlated part $F^{(d)}$ in our theoretical framework. 
For any polarization of the outgoing x-ray, the main feature C around 1-1.5 eV originates 
from the excitations from $e_g$ to $t_{2g}$ states. 
Particularly, $d_{yz}$ and $d_{xz}$ orbitals play a significant role for the main feature C. 
In contrast to $K$-edge RIXS, off-diagonal contributions are much more dominant than diagonal contributions, 
reflecting the multi-orbital nature of the $L$-edge RIXS process. 

\subsection{Predictions for lower-energy spectra}

\begin{figure}[htb]
\includegraphics[width=0.5\columnwidth]{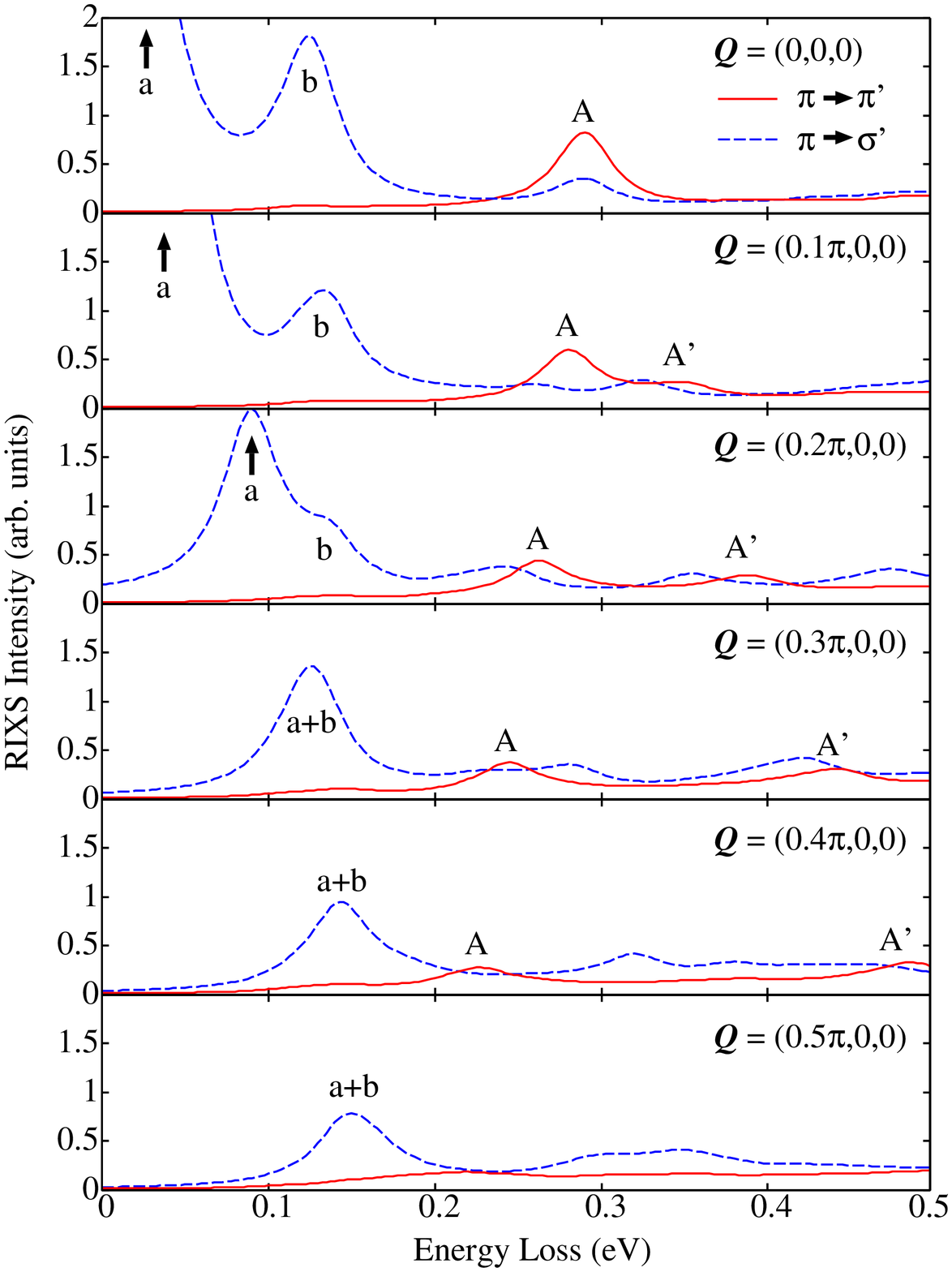}
\caption{(Color online)
Dependence of low-energy RIXS spectra on x-ray momentum transfer 
{\bf Q} along the [100] direction.}
\label{fig5}
\end{figure}

We turn our attention to the calculated results for lower excitation energies, below 0.5 eV. 
In Fig.~\ref{fig5}, we show the momentum dependence of calculated RIXS spectra 
along the symmetry line from ${\bf Q}=0$ to ${\bf Q}=(\pi/2, 0, 0)$. 
Our calculation suggests that the feature A exhibits a significant momentum dependence: 
A splits into the main low-energy feature A and higher-energy weak feature A', as ${\bf Q}$ goes away from ${\bf Q}=0$. 
Both of A and A' become diffusive at ${\bf Q}=(\pi/2, 0, 0)$. 
For lower excitation energies below 0.2 eV, we predict two significant peaks a and b, 
which are also maybe hidden by the elastic peak in the present experiment. 
While the feature b does not exhibit momentum dependence at all, 
the feature a shifts up to around 0.15 eV excitation energy, 
and merges to b around ${\bf Q}=(0.3\pi, 0, 0)$. 

All of the features A, a and b depend strongly on the polarization direction of outgoing x-rays. 
Particularly, a and b are predicted to vanish for the polarization condition $\pi \rightarrow \pi'$. 
As inferred from the difference in the polarization dependence, the low energy features, 
A, a and b substantially differ from the higher-energy features, B, C, and D. 
Our calculation indicates that the lower-energy features A, a and b have significant contribution not 
from spin-conserved excitations but from spin-flipped excitations. 

\begin{figure}[htb]
\includegraphics[width=0.5\columnwidth]{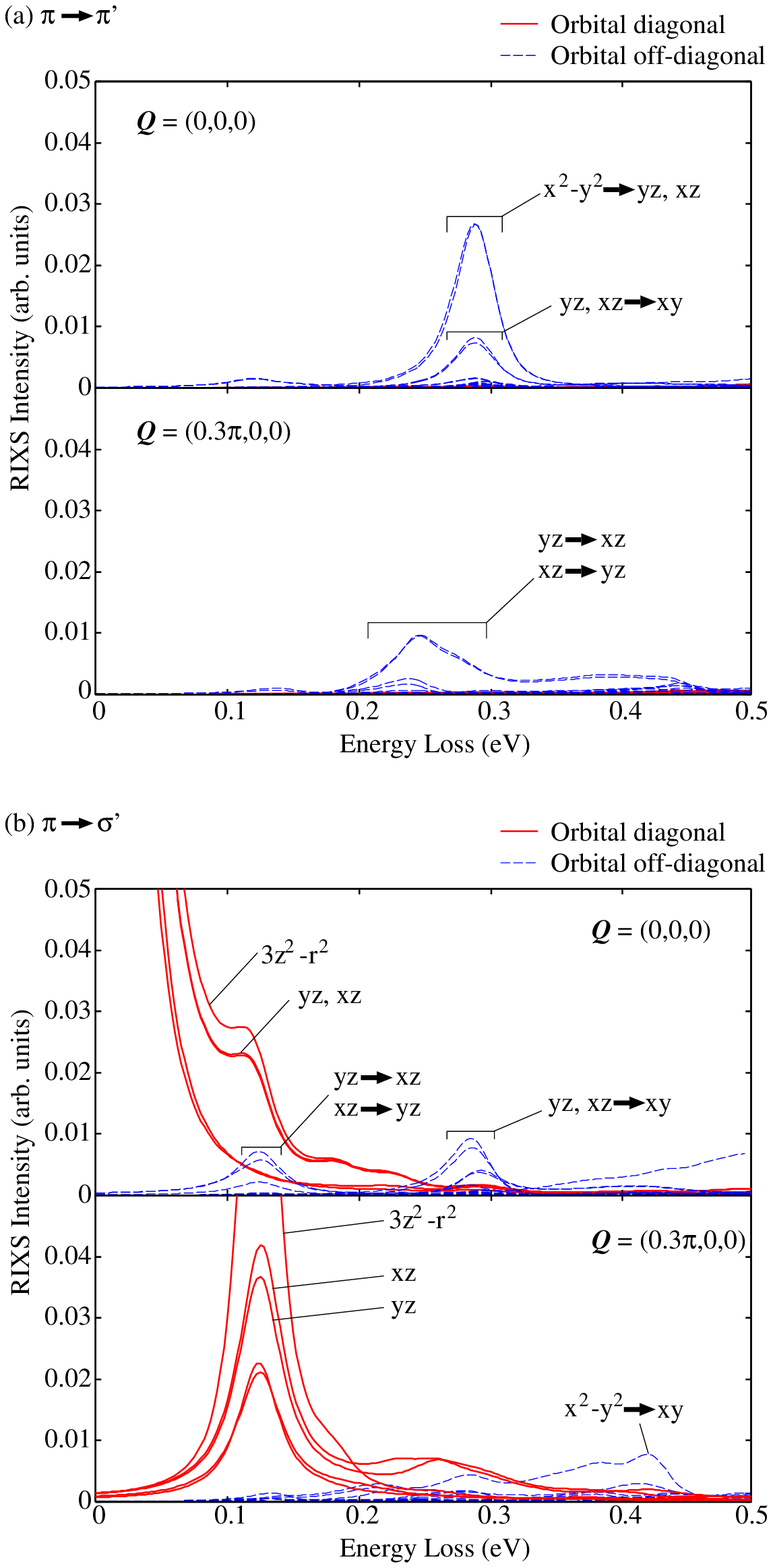}
\caption{(Color online)
Orbital-resolved intensities for spin-flipped (with respect to the [110] direction) channels. 
X-ray polarization conditions are (a) $\pi \rightarrow \pi'$ and (b) $\pi \rightarrow \sigma'$. 
The entire 25 processes including both diagonal and off-diagonal orbital excitations 
are plotted in each panel. 
Thick solid and thin broken curves represent diagonal and off-diagonal 
orbital excitations, respectively.}
\label{fig6}
\end{figure}

To study the origin of the low-energy features A, a and b, 
we show the orbital-resolved contributions in the spin-flipped (with respect to the [110] direction) channel in Fig.~\ref{fig6}. 
Clearly, contributions from the degenerate $yz$ and $xz$ orbitals become more significant 
below 0.5 eV, than above 0.5 eV. 
As shown in Fig.~\ref{fig6}, the feature A gains intensity from spin-flipped off-diagonal orbital 
excitations with predominant contribution from the $yz$ and $xz$ orbitals. 
At ${\bf Q}=0$, diagonal orbital excitations are almost irrelevant to the feature A 
in both polarization conditions. 
As the momentum transfer ${\bf Q}$ goes away from ${\bf Q}=0$, 
the excitation A starts gradually to be dominated by the $yz$ and $xz$ orbital character. 

The lower-energy features a and b arise 
from orbital-diagonal spin-flipped excitations, as shown in Fig.~\ref{fig6}(b). 
Although these are the same kind of single-magnon excitations as observed 
in copper oxides, spin-flipping is possible in each of the five $d$ orbitals in the iron pnictides. 
The strongest contribution is given by the $3z^2-r^2$ state, followed by the $yz$ and
$xz$ states.

The agreement between the experiment and the calculation for the higher-energy part of the spectrum above 0.5 eV lends support 
to the validity of the used electronic structure and subsequent theoretical treatment. 
However, an experimental confirmation of the predicted characteristic low-energy features A, a and b in the 1111 iron pnictide families 
using ultra-high energy resolution will be required to further validate our theoretical findings.

\section{Discussion}

In BaFe$_2$As$_2$ (122) systems~\cite{Zhou2012}, a dispersion of over 100 meV was observed 
for the spin excitation as a function of momentum transfer. 
Based on the similarity in the energy dispersion and excitation energy range,
the observed low-energy spin excitation may correspond to the feature a or b in our results.
However, we should note the feature A in our results has not been observed in the 122 systems. 
This is not surprising, since the electronic structure of 122 systems is sufficiently different from that of 1111 systems. 
In general, 122 systems are more three-dimensional than 1111 systems. 
In addition, the DOS of oxygen, absent in 122 systems, was observed near $E_F$ in PrFeAsO$_{1-\delta}$ by means of oxygen XAS and XES~\cite{Freelon2010}. 
Therefore the character of the Fe-$d$ orbitals involved in the excitation could also be different, which
crucially influences the RIXS response in intermediate excitation energies. 

The comparison with neutron scattering measurements provides an important perspective. If the spin-orbit coupling among the Fe-$d$ electrons is weak 
as in most of $3d$ transition-metal compounds, neutron scattering can only detect spin-flipped excitations without orbital change. Therefore the feature A, 
which is predominantly an orbital off-diagonal excitation, cannot effectively be observed in neutron scattering. On the other hand, features a and b, 
which predominantly originate from orbital-diagonal spin-flipped excitations, can be regarded as the counterparts of spin-wave modes observed in neutron scattering. 
In fact, the calculated excitation energies of the feature a ($\sim$ 150 meV at ${\bf Q}_{XY} = (0.5\pi,0)$) are consistent with those of the spin waves observed 
in 122 systems using neutron scattering~\cite{Dai2015}, bearing in mind that direct comparisons with 122 systems should be taken with caution, as mentioned above. 

We would like to compare our calculation with the precedent theoretical study 
by Kaneshita and collaborators~\cite{Kaneshita2011}. 
While similar low-energy features to the features a and b in our study are observed in their calculated results at a sight, 
substantial differences can actually be found, despite the fact that both studies determined the AF ground state within HFA and used the RPA. 
The most striking difference resides in the dominant orbital character at low energies: 
The $xy$ orbital plays a dominant role for low-energy spin excitations in their calculations, 
while it is the $3z^2-r^2$, $yz$ and $xz$ orbitals in ours. 
This contrast most likely arises from the difference in used electronic states. 
In their study, the degeneracy between the $yz$ and $xz$ orbitals seems to be lifted~\cite{Nomura2015c}.
Therefore we consider that their calculation holds for the orthorhombic cases, while our calculation is valid for the tetragonal cases.
Another noticeable difference is the intensity of magnon excitations around ${\bf Q}=0$. 
In their calculation, magnon intensity vanishes at ${\bf Q}=0$, 
while it is divergent in our calculation as shown in Fig.~\ref{fig5}. 
This difference originates in the different approximation used. 
The so-called fast-collision approximation (FCA) was used in their study, 
and consequently the RIXS intensity was expressed effectively using the imaginary part 
of the correlation function, while we did not adopt FCA in our theoretical framework. 
Recently, Igarashi and Nagao discussed effects beyond FCA and pointed out that symmetry breaking 
due to AF long-range ordering can cause such a kind of divergence in RIXS intensity toward ${\bf Q}=0$~\cite{Igarashi2015}. 
According to them, anisotropic terms which are not included in FCA considerably enhance the spin-flip RIXS intensity around ${\bf Q} = 0$, 
as seen in our calculated spectra. Whether such an effect of symmetry breaking is actually observed experimentally or not is an interesting issue. 
To settle it, we should measure both dependences on the polarization of the ingoing and outgoing x-rays with high energy-momentum resolution, 
which at present requires further advancements in experimental instrumentation. 

In our calculation, the long-range AF ordering was assumed to be fully developed. 
However, this does not imply that our calculation is applicable only to the AF ordering state 
below the N\'eel temperature. The RIXS process occurs on a femtosecond timescale, amply short compared with usual spin dynamics. 
In addition, the length scale, which can be estimated to be the product of the time scale 
and the velocity of electrons or elementary excitations, 
is very short compared with the usual AF magnetic correlation length. 
Therefore we may approximately regard spins as almost frozen with the magnetic configuration 
during the RIXS process, which can be treated within HFA. 
We can expect that our calculation becomes more valid at low temperatures 
where tetragonal symmetry is still maintained but the AF correlation has evolved into critically slow spin fluctuations 
and sufficiently long-range correlation length. 

Comparison with the copper oxides is quite illuminating. 
In the copper oxides, spin excitations, $dd$ orbital excitations, and charge-transfer excitations are observed separately 
in a different excitation energy range~\cite{Sala2011,Schlappa2012}. 
On the other hand, in the iron-pnictides, spin excitations and $dd$ orbital excitations are overlapped and entangled in the low-energy region. 
This notable contrast arises from the following differences in electronic states between these two kinds of compounds: 
Firstly, in the copper oxides, the Cu-$d_{x^2-y^2}$ level is partially filled and well separated from the other completely 
filled Cu-$d$ orbitals by crystalline-field energies which are larger than the spin excitation energy. 
Therefore interference between spin excitations and $dd$ excitations are suppressed. 
In contrast, in the iron-pnictides, all of the five Fe-$d$ states are partially filled and $dd$ orbital excitations are allowed 
even in the low-energy range where spin excitations occur. 
Secondly, $dd$ orbital excitations in the iron pnictides are rather broad, 
while they emerge as well-resolved sharp peaks in the copper oxides. 
As naturally understood, the sharpness of the $dd$ orbital excitations reflects the strength of localization of the $d$ electrons. 
In fact, the $dd$ excitations in the copper oxides have been well described by the single-ion crystal field model~\cite{Sala2011}. 
This suggests that the $dd$ excitations in the copper oxides are strongly localized. 
In contrast to the copper oxides, an itinerant description seems to be more appropriate for the 1111 iron-pnictide system. 
The itinerancy of the Fe-$d$ electrons in the iron pnictides has been argued based on discussions 
on the origin of the SDW ordering~\cite{Cvetkovic2009,Knolle2010,Zhang2010,Wysocki2011}. 
Our present study is consistent with those arguments, since the observed features B, C and D are broad $dd$ excitations 
which can be appropriately described by using an itinerant model and be regarded as Fe-$d$ interband transitions. 
Thus the multi-orbital nature and relatively strong itinerancy of the Fe-$d$ electrons are responsible 
for the entanglement and overlapping of the low-energy spin and orbital excitations in the iron pnictides. 

\section{Conclusion}

In summary, we reported Fe $L$-edge RIXS measurements on a typical iron pnictide, PrFeAsO$_{0.7}$. 
Well-resolved RIXS features were observed around 0.5, 1-1.5, 2-3 eV. 
The underlying excitation processes were investigated theoretically on the basis 
of first-principle electronic bands, and interpreted as orbital excitations among the Fe-$d$ orbitals. 
A low-energy 0.25 eV feature, likely hidden by the elastic tail in the experiment, 
was theoretically predicted and assigned to spin-flipped orbital excitations 
with strong $d_{yz}$ and $d_{xz}$ components. Consistency between the experiment and calculation confirmed 
that the Coulomb interaction among Fe-$d$ orbitals is moderately strong ($U \approx$ 3 eV) in this system. 
Furthermore, momentum dependence of the RIXS features below 0.5 eV was predicted, with remarkable splitting 
and merging of the lower-energy peaks in momentum space. 
Pending improvements in instrumental energy and momentum resolution in the next generation of RIXS spectrometers 
should enable experimental confirmation of these low-energy features and their behavior, 
which could not be observed in the present experiment. 
Finally, the contrast between these $L$-edge data and previously reported $K$-edge data 
on the same systems highlights the benefit of a combinatorial study using both edges; 
$K$ edge to probe diagonal $dd$ interband transitions and ligand-to-metal charge transfer, 
$L$ edge to probe off-diagonal $dd$ interband transitions and spin excitations.

\acknowledgements{
We would like to thank Dr. H. Gretarsson, Dr. E. Kaneshita, Prof. Y.J. Kim and Prof. T. Tohyama for invaluable communications. 
The synchrotron radiation experiments were performed at the BL07LSU beamline at SPring-8 with the approval 
of the Japan Synchrotron Radiation Research Institute (JASRI) (Proposal No. 2011B7420).}


\end{document}